\documentclass[aps,pra,twocolumn,preprintnumbers,amsmath,amssymb,floatfix,nofootinbib,reprint]{revtex4-1}

\usepackage{amsmath}
\usepackage{amsfonts}
\usepackage{amssymb}
\usepackage{graphicx}%
\usepackage{dcolumn}%
\usepackage{bm}%
\usepackage{color}
\usepackage{multirow}
\usepackage{hyperref}

\begin{document}

%
%

\title{X-ray assisted nuclear excitation by electron capture in optical laser-generated plasmas}

%
%

\author{Yuanbin \surname{Wu}}
\email{yuanbin.wu@mpi-hd.mpg.de}
\affiliation{Max-Planck-Institut f\"ur Kernphysik, Saupfercheckweg 1, D-69117 Heidelberg, Germany}

\author{Christoph H. \surname{Keitel}}
\affiliation{Max-Planck-Institut f\"ur Kernphysik, Saupfercheckweg 1, D-69117 Heidelberg, Germany}

\author{Adriana \surname{P\'alffy}}
\email{palffy@mpi-hd.mpg.de}
\affiliation{Max-Planck-Institut f\"ur Kernphysik, Saupfercheckweg 1, D-69117 Heidelberg, Germany}

\date{\today}

%
%
%
%
%
%
%
\begin{abstract}

X-ray assisted nuclear excitation by electron capture (NEEC) into inner-shell atomic holes in a plasma environment generated by strong optical lasers is investigated theoretically. The considered scenario involves the interaction of a strong optical laser with a solid-state nuclear target leading to the generation of a plasma. In addition, intense x-ray radiation from an X-ray Free Electron Laser (XFEL) produces inner-shell holes in the plasma ions, into which NEEC may occur.  As case study we consider the $4.85$-keV  transition starting from the 2.4 MeV long-lived $^{\mathrm{93m}}$Mo isomer  that can be used to release the energy stored in this metastable nuclear state.  We find that the recombination into $2p_{1/2}$ inner-shell holes is most efficient in driving the nuclear transition. Already at few hundred eV plasma temperature, the generation of inner-shell holes can allow optimal conditions for NEEC, otherwise reached for steady-state plasma conditions in thermodynamical equilibrium only at few keV. The combination of x-ray and optical lasers presents two advantages: first,  NEEC rates can be maximized at plasma temperatures where the photoexcitation rate remains low. Second, with mJ-class optical lasers and an XFEL repetition rate of $10$ kHz, the NEEC excitation number can reach $\sim 1$ depleted isomer per second and is competitive with  scenarios recently envisaged at petawatt-class lasers.

\end{abstract}

\maketitle

\section{Introduction}

Intense coherent light sources available today, covering a large frequency range from optical to x-ray light, open unprecedented possibilities for the field of laser-matter interactions \cite{DiPiazzaRMP2012}. In particular, not only do they address the electronic dynamics, but they might also influence the states of atomic nuclei. Novel X-ray sources as the X-ray Free Electron Laser (XFEL) can drive, for instance, low-energy electromagnetic transitions in nuclei \cite{Chumakov2018}. High-power optical laser systems with up to a few petawatt power are very efficient in generating plasma environments \cite{bookDieter} that host complex interactions between photons, electrons, ions, and the atomic nucleus. Nuclear excitation in hot plasmas generated by optical lasers has been the subject of numerous works \cite{HarstonPRC1999, Gosselin2004, Gosselin2007, Morel2004-local, Morel2007-local-Hg, Morel2010-nonlocal, Comet2015-NEET, KenL2000.PRL, Cowan2000.PRL, Gibbon2005.Book, Spohr2008.NJP, Mourou2011.S, U-experiment, NIF2018, Andreev2000, Andreev2001, Granja2007-list-nuclei, Renner2008, Gobet2011}. Also nuclear excitation in cold plasmas generated by XFELs \cite{Vinko2012.N} was shown to be relevant for a number of low-lying nuclear states  \cite{GunstPRL2014, GunstPOP2015}. Since plasmas in laser laboratories have temperatures restricted to few keV, the range of nuclear transitions that can be efficiently addressed is limited. As initial state, one can envisage either the nuclear ground state or   long-lived excited states, also known as nuclear isomers \cite{WalkerReview2016}. Isomers are particularly interesting due to their potential to store large amounts of energy over long periods of time \cite{WalkerN1999, AprahamianNP2005, BelicPRL1999, CollinsPRL1999, BelicPRC2002, CarrollLPL2004, PalffyPRL2007}. A typical example is $^{\mathrm{93m}}$Mo at $2.4$ MeV, for which an additional excitation of only $4.85$ keV could lead to the depletion of the isomer and release the stored energy on demand.

Nuclear excitation in plasmas may occur via several mechanisms. Apart from photoexcitation, the coupling to the atomic shell via processes such as nuclear excitation by electron capture (NEEC) or electron transition (NEET) \cite{GoldanskiiPLB1976, PalffyCP2010} may play an important role. In the resonant process of NEEC, a free electron recombines into an ion with the simultaneous excitation of the nucleus. The energy set free by the recombining electron has to be an exact match for the nuclear transition energy. Theoretical predictions have shown that as a secondary process in the plasma environment, NEEC may exceed the direct nuclear photoexcitation at the XFEL. For the $4.85$ keV transition above the $^{\mathrm{93m}}$Mo isomer, the secondary NEEC in the plasma  exceeds direct XFEL photoexcitation by approximately six orders of magnitude \cite{GunstPRL2014,GunstPOP2015}. Just recently, another theoretical work has shown that by tailoring optical-laser-generated plasmas to harness maximum nuclear excitation via NEEC, a further six orders of magnitude increase in the nuclear excitation and subsequent isomer depletion can be reached \cite{WuPRL2018, GunstPRE2018} compared to the case of cold XFEL-generated plasmas. 

\begin{figure}[h!]
   \includegraphics[width=\linewidth]{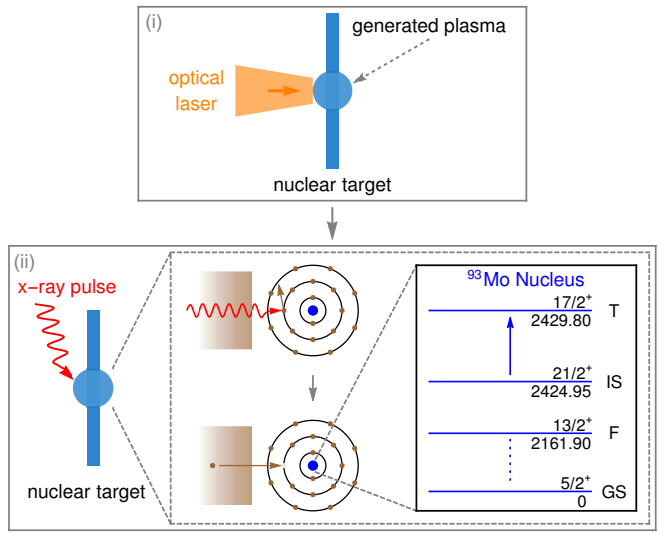}
    \caption{Sketch of x-ray assisted NEEC in an optical laser-generated plasma. In the first step (i), the optical laser interacts with a nuclear solid-state target generating a plasma. In the second step (ii), the XFEL interacts with the plasma to produce inner-shell holes into which NEEC can occur. The partial level scheme of $^{\mathrm{93m}}$Mo is shown in (ii). The nuclear isomeric (IS), triggering (T), intermediate (F), and the ground-state (GS) levels are labeled by their spin, parity, and energy in keV, taken from Ref.~\cite{ensdf-web}.} 
    \label{fig:setup}
\end{figure}

As a general feature of NEEC, capture is most efficient and the excitation cross sections are highest for electron recombination into inner-shell vacancies.   This holds true also for the   case of $^{\mathrm{93m}}$Mo, for which the largest NEEC cross section occurs for the capture into the $L$-shell vacancies, while capture into the inner-most $K$-shell is energetically forbidden \cite{GunstPRL2014,GunstPOP2015, WuPRL2018, GunstPRE2018}. However, in a plasma scenario it is difficult to simultaneously facilitate both the existence of inner-shell vacancies and the corresponding free electron energies that are allowed by NEEC. In order to decouple these two requirements,  in this work we consider a scenario in which a solid-state nuclear target interacts with both an optical and an x-ray laser. The optical laser generates the plasma environment and is responsible for the free electron energy distribution, while the x-ray laser interacts efficiently with inner-shell electrons producing the optimal atomic vacancies for NEEC. The sketch of this two-step scenario is illustrated in Fig.~\ref{fig:setup}. As a case study we consider again the $4.85$-keV nuclear transition starting from the long-lived excited state of $^{\mathrm{93m}}$Mo at $2.4$ MeV.  Apart from allowing for a practical isomer depletion scenario, $^{\mathrm{93m}}$Mo is  also interesting because of the recently reported observation of isomer depletion of the $4.85$-keV transition attributed to NEEC \cite{Chiara2018} and the subsequently raised question marks about the observed depletion mechanisms  \cite{WuPRL2019}.

Our results show that for low-temperature plasmas, the XFEL-generated inner-shell holes can substantially enhance the NEEC rates making them competitive with the case of high-temperature plasmas in thermodynamical equilibrium (TE). For low plasma temperatures, the x-ray assisted NEEC process is dominant. This holds true for electron densities in the range of $10^{19}-10^{21}$ cm$^{-3}$, while for high densities of $10^{23}$ cm$^{-3}$, recombination occurs too fast and substantially reduces the lifetimes of the inner-shell holes. The advantages of  x-ray-assisted NEEC in optical laser-generated plasma are two-fold. First, by requiring only a small plasma temperature, the competing nuclear photoexcitation rate is kept small and NEEC is dominant. Thus observation of photons belonging to the isomer decay cascade could be undoubtedly attributed to NEEC.  Second, the isomer depletion signal becomes detectable even when using a mJ optical laser, the excitation  being competitive with the predicted values for petawatt-class laser facilities. The requirements for optical laser power are therefore substantially lowered by the presence of the XFEL. Allegedly, XFEL facilities are at present more scarce than petawatt optical laser ones. However,  a combination of optical and X-ray lasers such as the Helmholtz  International Beamline for Extreme Fields HIBEF \cite{hibef-web} is already envisaged at the European XFEL \cite{europeanXFEL-web}, rendering possible in the near future the scenario investigated here.

The paper is structured as follows. In Sec.~\ref{sec:theory}, we introduce the scenario of x-ray-assisted NEEC in optical laser-generated plasmas and discuss the theoretical grounds of our calculations. Our numerical results for NEEC rates in  plasmas with XFEL-generated inner-shell holes and in particular for the two-laser setup are presented in Sec.~\ref{sec:numres}. The paper concludes with a brief summary in Sec.~\ref{sec:con}.

\section{Theoretical approach \label{sec:theory}}

In the following we investigate the scenario of a nuclear solid-state target interacting with a strong optical laser and an XFEL as depicted in Fig.~\ref{fig:setup}. 
High-power optical lasers are efficient in generating plasmas over a broad parameter region as far as both temperature and density are concerned. In turn, XFELs can interact with inner-shell electrons efficiently. The process is envisaged in two stages. At the plasma generation stage, the optical laser generates and heats the plasma and shortly after the end of the pulse we assume TE is reached. The mechanisms of plasma generation and heating are either the direct interaction of the laser with the target electrons for the low density case, or the secondary process of interaction of the hot electrons (directly heated by the laser) with the solid target for the high density case \cite{WuPRL2018, GunstPRE2018}. The time required to reach the TE steady state depends on the plasma density and temperature. According to the rate estimates based on the radiative-collisional code FLYCHK \cite{FLYCHK2005, FLYCHK-IAEA}, 
this varies from the order of $10$ fs for solid-state density to the order of $10$ ps for low density $\sim 10^{19}$ cm$^{-3}$ at temperature $\sim 1$ keV \cite{GunstPRE2018}. This timescale for reaching the TE steady state with regard to the atomic process is much shorter than the plasma lifetime estimated from the hydrodynamic expansion of the plasma for the considered conditions \cite{GunstPRE2018}.

In the second stage, the XFEL starts producing inner-shell holes via photoionization \cite{Vinko2012.N}. In this process, the actual number of x-ray photons in the XFEL pulse plays a more important role than the coherence properties of the latter. The energy of the photons is tuned to the ionization threshold energies for producing a hole in the $L$ shell. This energy is for all considered cases smaller than the  nuclear excitation energy, such that we do not expect any nuclear photoexcitation directly or via secondary photons from the XFEL interacting with the solid-state target.   
The XFEL pulse will additionally heat the plasma and shift the temperature by few tens of eV. However, since this shift is small compared to the plasma temperature induced by the optical laser, we  consider in first approximation for our calculation that the temperature is constant throughout the production of inner-shell holes. The additional plasma heating generated by the XFEL is expected to change the result on the level of a few percent. For the regions where only the XFEL hits the solid-state target, a cold plasma will be produced, but the corresponding contribution to the overall NEEC excitation is expected to be orders of magnitude smaller than from the optical-laser-generated plasma region.

Since the optical and x-ray lasers are supposed to act on the same target, it is important that some of their parameters such as intensity and repetition rate have similar orders of magnitude.  In Refs.~\cite{WuPRL2018, GunstPRE2018}, parameters of petawatt-class optical lasers were adopted for the calculations. However, the two types of facilities do not match well \cite{DansonHPLSE2014,eurXFELc-web}: 
 (i) Petawatt-class optical lasers are much stronger than XFELs (i.e., contain a larger number of photons); (ii) XFELs have higher repetition rates than petawatt-class lasers. For this reason in the present scenario we focus on the mJ-class optical lasers which could match XFELs in both  power and repetition rate.

Following the studies in Refs.~\cite{GunstPRL2014, GunstPOP2015, WuPRL2018, GunstPRE2018}, we consider the $4.85$-keV nuclear transition starting from a long-lived excited state of $^{\mathrm{93m}}$Mo at $2.4$ MeV, which has a half-life of $6.85$ h. The $^{\mathrm{93m}}$Mo isomer can be depleted by driving the $4.85$ keV electric quadrupole ($E2$) transition from the isomer to a triggering level (T), which subsequently decays via a cascade to the ground state. The partial level scheme of $^{\mathrm{93m}}$Mo is shown in Fig.~\ref{fig:setup}. The considered nuclear target is a Niobium target with a fraction of $^{\mathrm{93m}}$Mo isomer embedded in it. A $^{\mathrm{93m}}$Mo isomer fraction $f_{\rm{iso}} \approx 10^{-5}$ embedded in the Niobium target can be generated by intense ($\geqslant 10^{14}$ protons/s) proton beams \cite{GunstPRL2014} via the reaction $^{93}_{41}$Nb($p$, $n$)$^{\rm{93m}}_{42}$Mo \cite{exfor}.

\subsection{NEEC rates in plasmas \label{subsec:nrateh}}

The calculation of NEEC rates in plasma environments requires three important ingredients: the actual NEEC cross sections for capture into different atomic orbitals, the existence of atomic vacancies in the respective orbitals and the free electron distribution. At TE the NEEC rate, free electron flux and the charge state distribution can be written as a function of the plasma temperature. The NEEC rate in the plasma can be obtained by the summation over all charge states $q$ and all capture channels $\alpha_d$,
\begin{equation}
  \lambda_{\rm{neec}} (T_e, n_e) = \sum_{q, \alpha_d} P_q(T_e, n_e) \int dE    \sigma_{q}^{\alpha} (E)\phi_e(E, T_e, n_e),
\end{equation}
where $P_q$ is the probability to find ions of charge state $q$ in the plasma as a function of electron temperature $T_e$ and density $n_e$, $\sigma_{q}^{\alpha}$ is the NEEC cross section, and $\phi_e$ is the free-electron flux.

The microscopic NEEC cross sections are calculated following the formalism in Refs.~\cite{GunstPRL2014, GunstPOP2015, PalffyPRA2006, PalffyPRA2007}. The energy-dependent cross sections exhibit Lorentzian profiles centered on the discrete resonance energies which depend on the exact capture orbital,
\begin{equation}
  \sigma_{q}^{\alpha} (E) =  S_{q}^{\alpha}(E) \frac{\Gamma_{q, \alpha}/(2\pi)}{(E-E_{q, \alpha})^2 + \frac{1}{4} \Gamma_{q, \alpha}^2}\, ,
\end{equation}
where $S_{q}^{\alpha}(E)$ is the NEEC resonance strength, only slowly varying 
with respect to the electron energy,
and $E_{q, \alpha}$ and  $\Gamma_{q, \alpha}$ are the recombining electron energy and the natural width of the resonant state, respectively.  The resonant continuum electron energy $E_{q, \alpha}$ is given by the difference between the nuclear transition energy 4.85 keV and the electronic energy transferred to the bound atomic shell in the recombination process. For NEEC into the electronic ground state, the Lorentzian width is given by the nuclear state width and is  $10^{-7}$ eV for the 2429.80 keV level $T$ above the isomer \cite{ensdf-web}. The Lorentz profile can then be approximated by a Dirac-delta function. However, if the electron recombination occurs into an excited electronic configuration, the width of the Lorentz profile is determined by the electronic width, typically on the order of 1 eV \cite{CampbellADNDT2001}. This  value is still  small compared to the continuum electron energies of  few keV.

 The resonance strength $S_{q}^{\alpha}(E)$ depends on the matrix elements of the Hamiltonian coupling the electronic and nuclear degrees of freedom, which can be separated into an electronic and a nuclear part. The occurring nuclear matrix elements can be related to the reduced transition probability $B(E_2)$ for which the calculated value of $3.5$ W.u. (Weisskopf units) \cite{Hasegawa2011.PLB} has been adopted. The electronic matrix element is calculated  using relativistic electronic wave functions. The bound  atomic wave functions are obtained from the multi-configurational Dirac-Fock method implemented in the GRASP92 package \cite{Grasp92}, while for the continuum wave functions  solutions of the Dirac equation with $Z_{\rm{eff}} = q$ were used. Following the analysis in Ref.~\cite{GunstPRE2018}, we do not consider for the electronic wave functions the effects of the plasma temperature and density, which are sufficiently small to be neglected in the final result of the nuclear excitation. 
 
The free-electron flux $\phi_e(E, T_e, n_e)$ is obtained by modelling the  plasma by a relativistic Fermi-Dirac distribution. This approximation based on TE
is appropriate starting shortly after the end of the optical laser pulse, since thermalization occurs on a much shorter timescale much than the plasma lifetime over which NEEC takes place \cite{GunstPRE2018}. As already mentioned, we neglect here the shift in electron temperature induced by the XFEL which is expected to be few tens of eV for the case under consideration.

The  charge state distribution $P_q(T_e, n_e)$ after the optical laser pulse is computed using the radiative-collisional code FLYCHK \cite{FLYCHK2005, FLYCHK-IAEA} assuming the plasma to be in its nonlocal TE steady state. For the case of x-ray assisted NEEC, we additionally assume the presence of inner-shell holes in  the $2s$ and $2p$ orbitals of Mo. We note that FLYCHK uses scaled hydrogenic wave functions, which is expected to lead to errors for cold plasmas with constituents of mid- and high atomic number $Z$. However, the comparison with experimental data in Ref.~\cite{FLYCHK2005} shows that for the region of interest for our study, the  deviations of the charge-state distributions are on the level of $10$\%.

\subsection{Number of depleted isomers \label{subsec:nexc}}

In order to obtain the NEEC excitation number $N_{\rm{exc}}$, we integrate the NEEC rate $\lambda_{\rm{neec}}$ over the plasma volume $V_p$ and the approximate plasma lifetime. In the present work, we assume in a first approximation homogeneous plasma conditions (density, temperature, and charge-state distribution) over the plasma lifetime $\tau_p$. This assumption will most likely not be fulfilled in realistic laser-generated plasmas. However,  
the comparison with the more detailed hydrodynamic expansion model in Ref.~\cite{GunstPRE2018} has shown that this lifetime approximation provides reasonable NEEC results, with  a  $\sim 50\%$ of deviation for the case of low temperatures and $\sim 5\%$ for high temperatures ($\geqslant 6$ keV). In addition, our calculations here predict that both the NEEC rate and the resulting number of excited nuclei have a smooth dependence on density and temperature. We therefore expect that realistic spatial density and temperature gradients  will not change the order of magnitude of our results.

For homogeneous plasma conditions, the total number of excited nuclei can be written as
\begin{equation}
  N_{\rm{exc}} = N_{\rm{iso}} \lambda_{\rm{neec}}^{\rm{TE}} \tau_p + N_{\rm{iso}}^{h} \lambda_{\rm{neec}}^h \tau_h,
  \label{eq:nexc}
\end{equation}
where $N_{\rm{iso}}$ is the number of isomers in the plasma, and $\lambda_{\rm{neec}}^{\rm{TE}}$ is the total NEEC rate assuming the plasma is in TE steady state \cite{WuPRL2018, GunstPRE2018}. Here, $N_{\rm{iso}}^{h}$ is the number of isomers with an x-ray generated inner-shell $h$ hole in the atomic shell ($h$: $2s$, $2p_{1/2}$, or $2p_{3/2}$) in the plasma, $\lambda_{\rm{neec}}^h$ is the NEEC rate for the capture into the inner-shell $h$ hole, and $\tau_h$ is the lifetime of the $h$ hole. The number of isomers can be estimated by $N_{\rm{iso}} = f_{\rm{iso}} n_i V_p$, where $n_i$ is the ion number density in the plasma. Assuming a spherical plasma, the plasma lifetime is approximatively given by \cite{GunstPOP2015, WuPRL2018, GunstPRE2018, KrainovS2002}
\begin{equation}
  \tau_p = R_p \sqrt{m_i/(T_e \bar{Z})},
\end{equation}
where $m_i$ is the ion mass, $\bar{Z}$ is the average charge state, and $R_p$ is the plasma radius. 

We now proceed to estimate the number of isomers $N_{\rm{iso}}^{h}$ with a hole in the $L$ shell. To this end we use the photon absorption cross section values from Ref.~\cite{EPDLa1989} considered at the ionization threshold energy. Depending on the particular charge state, the latter spans between approx. $2$ keV and $4.8$ keV for generating holes in the three orbitals of the $L$ shell. Over this interval, the absorption cross sections have a smooth dependence on the incoming photon energy and in particular vary insignificantly over the XFEL pulse energy width of approx. 20 eV \cite{europeanXFEL-web}.
For an order of magnitude estimate we  neglect the complicated XFEL pulse coherence properties and consider just the total x-ray photon number of approx. $10^{12}$ photons/pulse \cite{europeanXFEL-web}. The corresponding flux of incoherent photons together with the absorption cross section values that can reach a few times of $10^5$ barns for the considered $2s$ and $2p$ orbitals \cite{EPDLa1989} result in a hole production fraction of approx. 70$\%$ for the $2p_{3/2}$ hole and slightly smaller for the other two orbitals. It is likely that by taking into account the effect of the XFEL coherence properties on photoionization in a similar manner as in Refs.~\cite{CavalettoPRA2012, OreshkinaPRL2014}, these fractions of plasma ions with $L$-shell holes will increase. In the following, for an order of magnitude estimate of the isomer depletion in the x-ray assisted NEEC in optical-laser-generated plasmas, we will assume  that $N_{\rm{iso}}^{h} = N_{\rm{iso}}$. 

The hole lifetime $\tau_h$ is estimated based on the timescale $\tau_h^r$ of recombination of the generated hole computed with the FLYCHK code \cite{FLYCHK2005, FLYCHK-IAEA}, i.e., $\tau_h = \tau_{\rm{xfel}}$ if $\tau_h^r < \tau_{\rm{xfel}}$, $\tau_h = \tau_h^r$ if $\tau_{\rm{xfel}}< \tau_h^r < \tau_p$, and $\tau_h = \tau_p$ if $\tau_h^r > \tau_p$. Here, $\tau_{\rm{xfel}}$ is the duration of XFEL pulse. In the following we assume an XFEL pulse duration of 100 fs \cite{GunstPOP2015, europeanXFEL-web, eurXFELc-web}. We note that the plasma lifetime $\tau_p$ is on the order of $10$ ps or even longer for the plasma conditions under consideration. Thus the XFEL pulse duration is a few orders of magnitude shorter than the plasma lifetime.

\section{Numerical results \label{sec:numres}}

\begin{figure}[h!]
   \includegraphics[width=\linewidth]{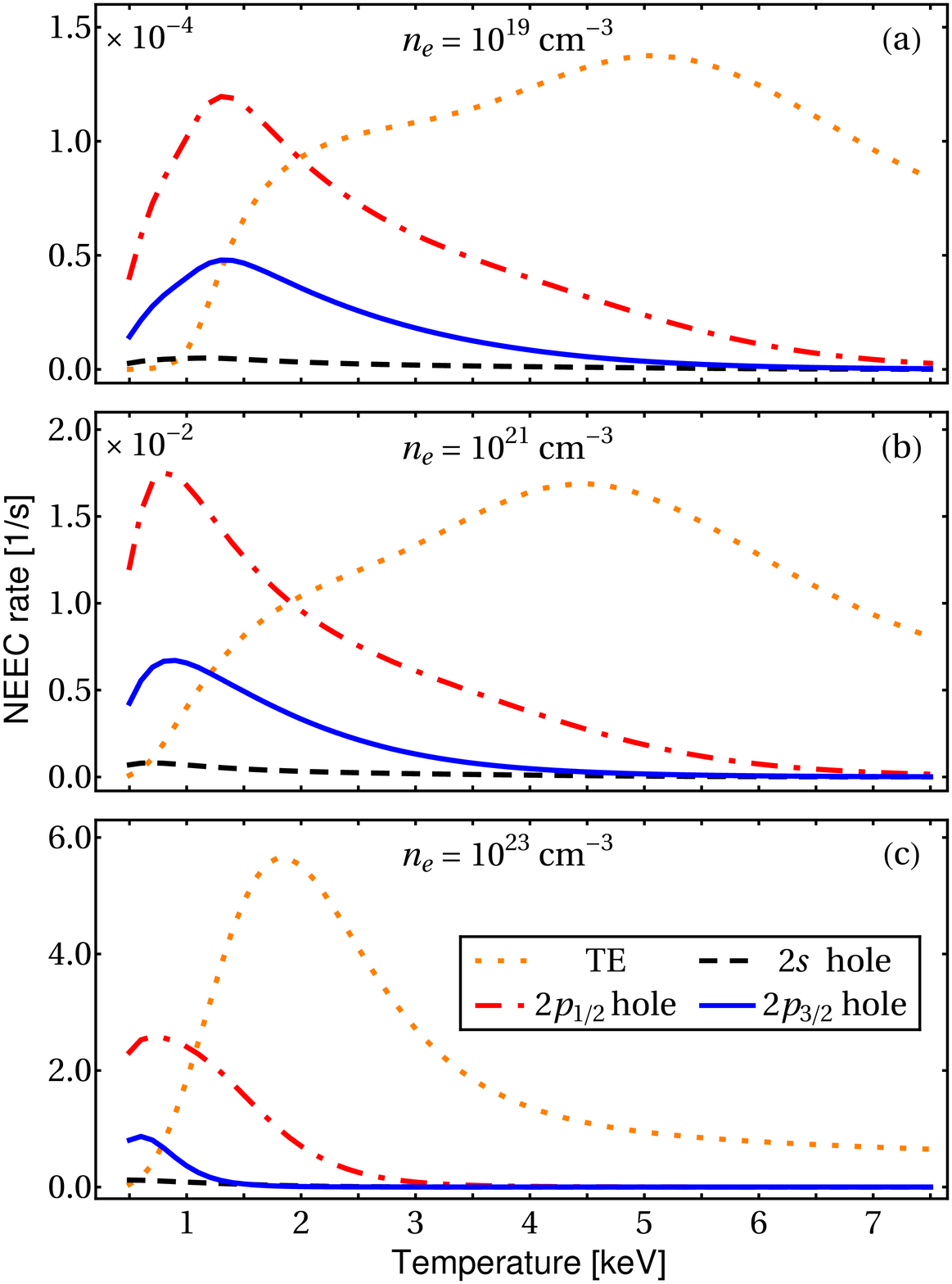}
    \caption{NEEC rates as function of plasma temperature, for selected electron densities (a) $n_e = 10^{19}$ cm$^{-3}$, (b) $n_e = 10^{21}$ cm$^{-3}$, and (c) $n_e = 10^{23}$ cm$^{-3}$.  NEEC occurs considering the TE steady state of the optical-laser generated plasma only (orange dotted curve) or in the x-ray assisted process by capture into an inner-shell $2s$   (black dashed curve), $2p_{1/2}$ (red dash-dotted) or $2p_{3/2}$ hole (blue solid curve). }
    \label{fig:neechrate}
\end{figure}

We calculate NEEC rates as function of plasma temperature for selected electron densities, for both TE conditions as previously discussed in Refs.~\cite{WuPRL2018, GunstPRE2018} and for capture into vacant inner-shell holes in the x-ray assisted process. For the latter we consider in first approximation the TE free electron flux and charge distribution with an additional inner-shell hole as the capture state. For the TE case, we consider $333$ NEEC capture channels \cite{WuPRL2018, GunstPRE2018}. The main capture channels involve recombination into the $L$ shell and $M$ shell, depending on  density and temperature \cite{WuPRL2018, GunstPRE2018}. The NEEC resonance strengths $S_{q}^{\alpha}(E)$ at the resonance energies for capture into $L$ shell lie between approx. $10^{-3}$ b eV and $10^{-5}$ b eV, while the NEEC resonance strength for capture into other shells is $\lesssim 10^{-5}$ b eV \cite{GunstPRE2018}. The resonance energy for  $L$-shell capture requires free electron energies between $52$ eV and $597$ eV  \cite{GunstPRE2018}, which should be well available at plasma temperatures of a few hundred eV. However, with such low temperatures, under the TE steady state $L$-shell vacancies are inexistent, thus allowing only for the capture into $M$ shell or further outer shells \cite{WuPRL2018, GunstPRE2018}. The inner-shell hole created by the XFEL could optimise the NEEC by decoupling the free electron energy condition from the capture state generation condition.
The results for NEEC rates in the TE and the x-ray assisted scenarios are presented and compared in  Fig.~\ref{fig:neechrate}. The capture into a $2p_{1/2}$ hole is the most efficient one, and the maximal NEEC rate for the capture into the $2p_{1/2}$ hole is comparable with the maximal total NEEC rate for plasmas in the TE steady state. As an advantage, the optimal conditions for  NEEC into the inner-shell hole occur at much lower temperatures than for the case of TE conditions where the temperature governs the charge state distribution and population of the capture orbitals.

\begin{figure}[h!]
   \includegraphics[width=\linewidth]{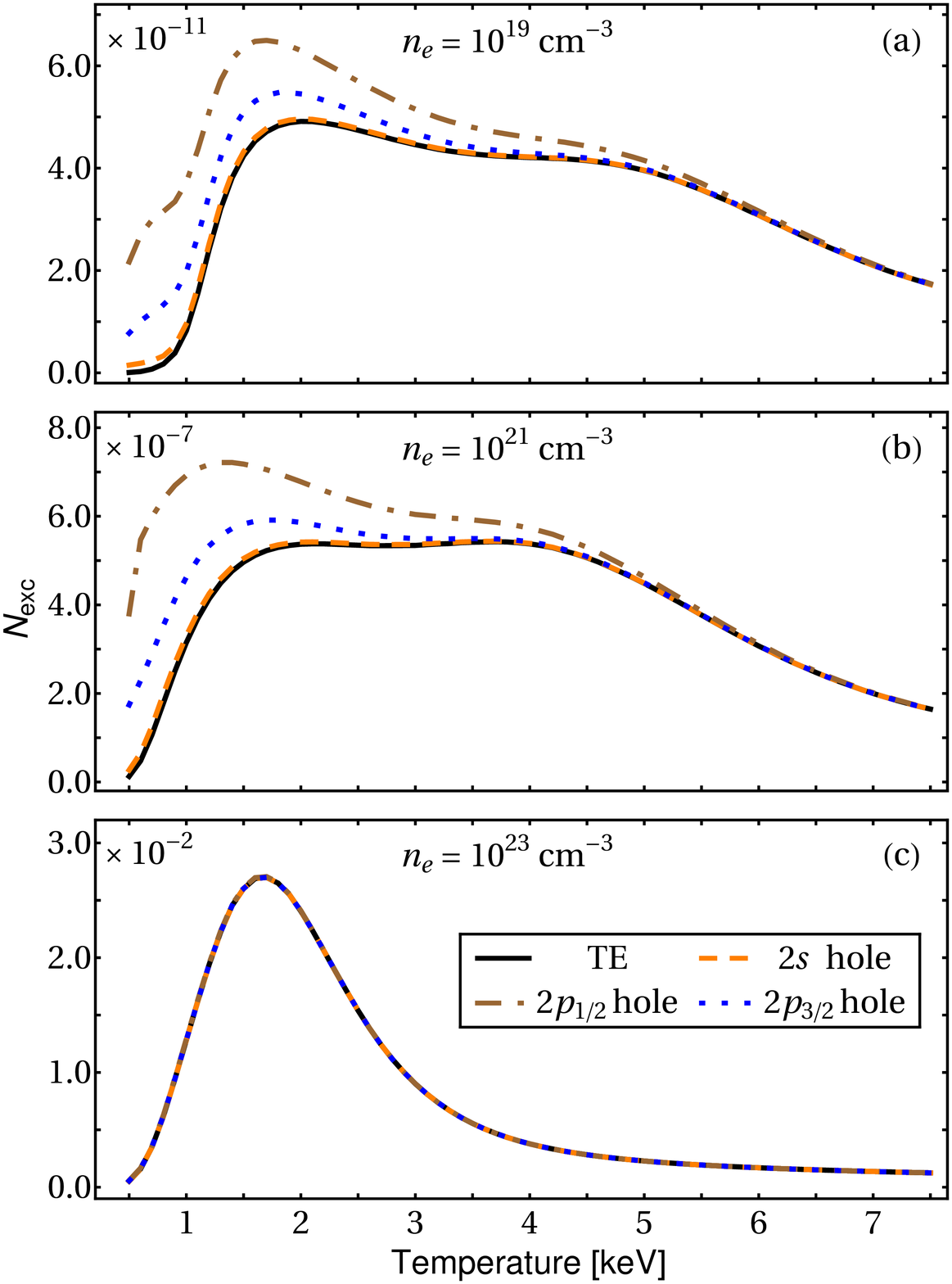}
    \caption{NEEC excitation number $N_{\rm{exc}}$ in spherical plasmas as function of plasma temperature for selected electron densities (a) $n_e = 10^{19}$ cm$^{-3}$, (b) $n_e = 10^{21}$ cm$^{-3}$, and (c) $n_e = 10^{23}$ cm$^{-3}$. The considered spherical plasma radius is $10$ $\mu$m. NEEC occurs considering the TE steady state of the optical-laser generated plasma only (black solid line) or in the x-ray assisted process by capture into an inner-shell $2s$   (orange dashed line), $2p_{1/2}$ (brown dash-dotted line) or $2p_{3/2}$ hole (blue dotted line).
      }
    \label{fig:sphn}
\end{figure}

We proceed to calculate the number of depleted isomers in Eq.~(\ref{eq:nexc}) considering a spherical plasma  of $10$ $\mu$m radius. The results are shown in Fig.~\ref{fig:sphn}. For 
the low-density cases $n_e = 10^{19}$ cm$^{-3}$ and $n_e = 10^{21}$ cm$^{-3}$, 
the presence of an XFEL-produced inner-shell hole leads to a great enhancement of the excitation number $N_{\rm{exc}}$ compared to TE conditions.  In the x-ray assisted scenario, already for small plasma temperatures we obtain excitation values which are comparable with the optimal $N_{\rm{exc}}$ in the TE plasma steady state which occurs only at temperatures of a few keV. This lowers the requirements on the optical laser power that generates and heats the plasma. 

For higher densities, the effects of the inner-shell hole become negligible. This happens because for high densities, the electron recombination (via other processes than NEEC) becomes very fast, such that the lifetime of the inner-shell hole $\tau_h$ is significantly decreased. This reduces the importance of the second term in the sum of Eq.~(\ref{eq:nexc}) such that the total excitation is determined predominantly by the TE term. We note that the timescale of  recombination of the generated hole through atomic processes varies from the order of $10$ fs for solid-state density to the order of $10$ ps for low density, and the plasma lifetime $\tau_p$ is on the order of $10$ ps or even longer for the plasma conditions under consideration. Thus for high density case, the lifetime $\tau_h$ of the inner-shell hole is determined by the XFEL duration, which is also a few orders of magnitude shorter than the plasma lifetime as discussed in Sec.~\ref{subsec:nexc}.

 As discussed in Refs.~\cite{WuPRL2018, GunstPRE2018}, nuclear photoexcitation from the thermal photons in the plasma dominates over NEEC starting with temperatures of a few keV ($T_e > 1.6$ keV for the density of $n_e = 10^{21}$ cm$^{-3}$). While photoexcitation enhances the total number of depleted isomers, it does not allow for an unambiguous identification of the nuclear excitation mechanism that led to isomer depletion. Thus, the x-ray assisted process which leads to an enhancement of the excitation at low temperatures, where photoexcitation is not yet dominant, may be helpful to clearly identify the nuclear excitation mechanism in the plasma.


\begin{figure}[h!]
   \includegraphics[width=\linewidth]{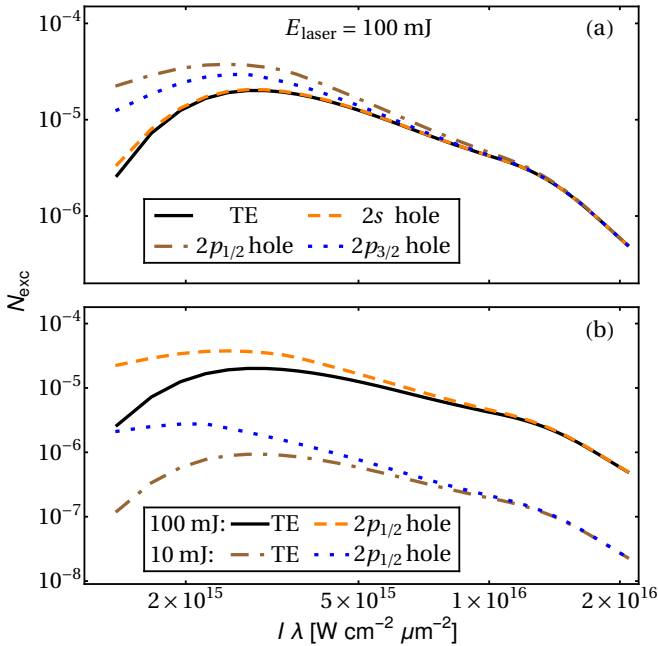}
    \caption{  $N_{\rm{exc}}$ as  function of the laser parameter product $I \lambda$. The electron density of the plasma is $n_e = 10^{21}$ cm$^{-3}$. (a) The results for a laser pulse with energy $E_{\rm{laser}} = 100$ mJ. NEEC occurs considering the TE steady state of the optical-laser generated plasma only (black solid line) or in the x-ray assisted process by capture into an inner-shell $2s$   (orange dashed line), $2p_{1/2}$ (brown dash-dotted line) or $2p_{3/2}$ hole (blue dotted line). 
   (b) comparison of TE and $2p_{1/2}$ inner-shell hole cases for a 100 mJ laser (black solid line and orange dashed line, respectively) and a 10 mJ laser (brown dash-dotted line and blue dotted line, respectively).}
    \label{fig:scaln}
\end{figure}

In the following, we proceed to determine the required optical laser parameters to generate the plasma conditions optimal for  NEEC. We focus on the case of low-density plasma since it is here that the x-ray assisted scenario presents its advantages.  In the case of low density (underdense) plasma, the plasma generation can be modelled with the help of the so-called scaling law, following the discussion in Refs.~\cite{WuPRL2018, GunstPRE2018}. We adopt here the widely used ponderomotive scaling law in the nonrelativistic limit \cite{Brunel1987, BonnaudLPB1991, GibbonPPCF1996},
\begin{equation}
  T_e \approx 3.6 I_{16} \lambda_{\mu}^2 ~ {\rm{keV}},
\end{equation}
to connect the electron temperature to the laser parameters. Here $I_{16}$ is the laser intensity in units of $10^{16}$ W/cm$^2$ and $\lambda_{\mu}$ is the wavelength in microns. The total number of electrons in the plasma can be estimated by the absorbed laser energy,
\begin{equation}
  N_e = f_{\rm{abs}} E_{\rm{laser}}/ T_e,
\end{equation}
where $f_{\rm{abs}}$ is the laser absorption coefficient, and $E_{\rm{laser}}$ is the energy of the laser pulse. Then the realistic plasma volume can be found by $V_p = N_e/n_e$.  Since experimental results in Refs.~\cite{PingPRL2014, PricePRL1995} show that the laser absorption is almost independent of the target material and thickness, we adopt here a universal absorption coefficient, which is an interpolation to theoretical results based on a Vlasov-Fokker-Planck code presented in Ref.~\cite{PingPRL2014}. For the considered intensity range of interest, the laser absorption coefficient $f_{\rm{abs}}$ lies between $0.1$ and $0.2$ \cite{WuPRL2018, GunstPRE2018}.

We focus on mJ-class optical lasers and consider two cases with $100$ mJ and $10$ mJ energies, respectively. For the electron density we adopt the value of  $n_e = 10^{21}$ cm$^{-3}$. The NEEC excitation numbers $N_{\rm{exc}}$ as function of the laser parameter product $I \lambda$ are shown in Fig.~\ref{fig:scaln} for both TE conditions and the x-ray assisted scenario. Our results show that for 
 low laser intensity,  $N_{\rm{exc}}$ is  enhanced by almost one order of magnitude compared to the case of the plasma in TE steady state. Assuming a repetition rate of $10$ kHz, which is in principle available for both mJ-class optical lasers and XFELs, the NEEC excitation number can reach as much as $\sim 1$ per second. This is at the same level as the one we can obtain from  plasmas generated by petawatt-class lasers \cite{WuPRL2018, GunstPRE2018}, with the advantage that mJ laser facilities have much better availability. 
 
 The experimental signature of the nuclear excitation would be a $\gamma$-ray photon of approximately $1$ MeV released in the decay cascade of the triggering level in $^{93}$Mo. Evaluations of the plasma blackbody and bremsstrahlung radiation spectra at this photon energy show that the signal-to-background ratio is very high \cite{WuPRL2018, GunstPRE2018}. As the emitted $\gamma$-rays are isotropic, ideally the  detector should cover a large solid angle. In this case, as the NEEC excitation numbers can reach $\sim 1$ per second, we expect that with a mJ-class optical laser and XFEL with a repetition rate of $10$ kHz,  thousands of counts could be collected per hour. 
 
At the European XFEL \cite{europeanXFEL-web} a combination of optical and XFEL lasers the will be soon available under the user consortium HIBEF \cite{hibef-web}.
 A commercial terawatt laser has been already installed  at the European XFEL, and a high energy ($100$ J per pulse) laser is further planned. HIBEF may soon provide the chance to investigate experimentally in the near future the scenario studied here. Alternatively, other XFEL facilities already operational or at present under construction   such as LCLS (LCLS II) \cite{LCLS-web}, SACLA \cite{SACLA-web}, and SwissFEL \cite{SwissFEL-web} in combination with commercially available mJ-class optical lasers  \cite{BackusRSI1998, NolteBook2016}, will open new opportunities to explore the role of inner-shell holes for NEEC. Furthermore, since the coherence properties of the x-ray source are less relevant for photoionization than for instance for x-ray quantum optics applications,  one can also envisage the use of  the optical-laser-driven x-ray sources \cite{FuchsNP2009, KneipNP2010, PhuocNP2012, PowersNP2014, CalendronHPLSE2018} as the one under construction at the Extreme Light Infrastructure (ELI) \cite{ELI-web} as a further laser facility option for investigating the x-ray assisted NEEC in plasma scenario.

\section{Conclusion \label{sec:con}}

We have investigated theoretically the scenario of x-ray assisted NEEC in the plasmas generated by mJ optical lasers. The role of the x-ray photons from an XFEL pulse is to produce $L$-shell holes which are available for electron recombination and therefore facilitate NEEC. 
 Our results show that capture into a $2p_{1/2}$ inner-shell hole is the most efficient one, and the maximal NEEC rate here is comparable with the maximal total NEEC rate for plasmas in TE steady state. The advantage is however that the optimal condition for the NEEC into the inner-shell hole happens at temperatures of a few hundred eV, much lower than the few keV defining the optimal condition for the total NEEC rate for plasmas in TE steady state. The enhancement at low temperatures (few hundred eV) is helpful to avoid nuclear photoexcitation from the thermal photons in the plasma discussed in Refs.~\cite{WuPRL2018, GunstPRE2018}, possibly leading to a clear NEEC signal. Our results also show that the  effect of inner-shell holes is negligible for the high density case ($n_e = 10^{23}$ cm$^{-3}$), as the recombination of the inner-shell hole is here too fast. Considering mJ-class optical lasers and XFELs with a repetition rate of $10$ kHz, the NEEC excitation number can reach $\sim 1$ per second, which is competitive with the NEEC in the plasmas generated by petawatt-class lasers \cite{WuPRL2018, GunstPRE2018}. Our results would be interesting for the envisaged facilities with a combination of optical and X-ray lasers such as HIBEF \cite{hibef-web} at the European XFEL \cite{europeanXFEL-web}.

\begin{acknowledgments}
The authors gratefully acknowledge fruitful discussions with J. Gunst.
\end{acknowledgments}

\bibliography{lxneecrefs}

\end{document}